\documentclass[aps,pra,eprint,superscriptaddress]{revtex4-2}

\usepackage{graphicx}% Include figure files
\usepackage{dcolumn}% Align table columns on decimal point
\usepackage{bm}% bold math
\usepackage{xcolor}
\usepackage{graphicx}
\usepackage{caption}
\usepackage{hyperref}
\hypersetup{colorlinks=true}

\begin{document}

\title{Experimental realization of para-particle oscillators}

\author{C. Huerta Alderete}
\email[e-mail: ]{aldehuer@umd.edu}
\affiliation{Joint Quantum Institute, Department of Physics, University of Maryland, College Park, MD 20742, USA}
\author{Alaina M. Green}
\affiliation{Joint Quantum Institute, Department of Physics, University of Maryland, College Park, MD 20742, USA}
\author{Nhung H. Nguyen}
\affiliation{Joint Quantum Institute, Department of Physics, University of Maryland, College Park, MD 20742, USA}
\author{Yingyue Zhu}
\affiliation{Joint Quantum Institute, Department of Physics, University of Maryland, College Park, MD 20742, USA}
\author{B. M. Rodr\'iguez-Lara}
\affiliation{Tecnologico de Monterrey, Escuela de Ingenier\'ia y Ciencias, Ave. Eugenio Garza Sada 2501, Monterrey, N. L., Mexico, 64849}
\author{Norbert M. Linke}
\affiliation{Joint Quantum Institute, Department of Physics, University of Maryland, College Park, MD 20742, USA}

\date{\today}
\begin{abstract}
    Para-particles are fascinating because they are neither bosons nor fermions.
    While unlikely to be found in nature, they might represent accurate descriptions of physical phenomena like topological phases of matter.
    We report the quantum simulation of para-particle oscillators by tailoring the native couplings of two orthogonal motional modes of a trapped ion.
    Our system reproduces the dynamics of para-bosons and para-fermions of even order very accurately.
    These results represent the first experimental analogy of para-particle dynamics in any physical system and demonstrate full control of para-particle oscillators.
  % \\ {\bf One-sentence summary: We report the first quantum simulation of para-particle oscillators realized by tailoring the coupling between the spin and two orthogonal motional modes of a trapped ion to reproduce the dynamics of para-bosons and para-fermions of even order very accurately.} 
\end{abstract}

%
%\keywords{}

\maketitle

%%%--------------------------------
\section{Introduction}
%%%--------------------------------
Investigating particles that are neither bosons or fermions is of fundamental interest in physics.
One example is the so-called para-particles, which were formally introduced in the early years of quantum mechanics as a generalization of bosons and fermions \cite{Green1953p270,Greenberg1965pB1155,Plyushchay1997p619}.
They fall into two categories, para-bosons and para-fermions, and are further classified by a parameter of deformation, or order of para-quantization, $p\geq 1$. 
Like ordinary particles, para-particles are characterized by their spin and dimensionality in Hilbert space; para-bosons have integer spin and an infinite-dimensional representation, while para-fermions have half-integer spin and a finite-dimensional representation. 
Standard bosons can be thought of as para-particles of order $p=1$.

During the initial development of the theory of para-particles, there was great interest in finding out if there are matter/field candidates for them in nature.
However, it was shown that it is not possible to distinguish a para-particle from a collection of standard particles by measurement \cite{Hartle1969p2043}.
As a result, no particle from the standard model is currently considered a para-particle candidate \cite{Greenberg1965pB1155,Baker2015p929}.
The field has become purely theoretical but still has produced a string of remarkable results. 
Para-bosons and para-fermions were considered as candidates for the particles of dark matter/dark energy \cite{Ebadi2013p057,Nelson2016p034039,Kitabayashi2018p043504} and as a tool to describe excitations in solids \cite{Safonov1991p109}.
In para-quark models quarks were considered as a para-fermion of order $p=3$ before the color model was introduced \cite{Bracken1973p1784}.
Proposals for the quantum simulation of para-bosons \cite{HuertaAlderete2017p013820} and para-fermions \cite{HuertaAlderete2018p11572} renewed the interest of the community in uncovering potential uses of para-particles, mainly from the vantage of statistical thermodynamics \cite{Hama1992p149,Stoilva2020p126421}, and their utility in the characterization of non-classical properties of light \cite{HuertaAlderete2017p043835,Mojaveri2018p529,Mojaveri2018p346,Mojaveri2018p1850134,Wei_Min2001p283} as well as possible applications in optics \cite{RodriguezWalton2020p043840}.
Despite this wave of new activity, to the best our knowledge, para-particle dynamics have not been realized experimentally, nor have the many theoretical predictions associated with them. 
Many physical systems have been put forward to produce experimental evidence, such as laser-inscribed arrays of evanescently coupled waveguides \cite{RodriguezWalton2020p043840} or hybrid systems involving two transmission line resonators controlled by a superconducting qubit \cite{Li2012p014303,Ma2014p062342,Strauch2010p050501}. 
Following the path laid out in Refs. \cite{HuertaAlderete2017p013820,HuertaAlderete2018p11572}, a quantum system involving a spin$-1/2$ degree of freedom coupled to two bosonic modes can be engineered to simulate para-bosons and para-fermions \cite{Chilingaryan2015p245501, HuertaAlderete2016p414001}. 
We choose a trapped ion system \cite{Liebfried2003p281,Haffner2008p155,Blatt2008p1476,Monroe2021p025001} for this experimental realization for its track record of motional state control \cite{Meekhof1996p1796,An2015p193,Um2016p11410,Debnath2018p073001,Jost2009p683,Ohira2019p060301}, tailored spin-motion interactions \cite{Chen2021p060311,Lemmer2018p073002}, and high fidelity operations including reliable state preparation and measurement. %[cite Oxford]. 
Trapped ions provide a highly controllable quantum environment, which grants access to phenomena in regimes that are not otherwise accessible in nature \cite{Schutzhold2007p201301,Lv2018p021027}. 

In this work, we report the analog quantum simulation of both kinds of para-particle oscillators by using the spin of a trapped atomic ion and two of its bosonic modes of motion in the trap, tailoring laser-induced couplings between them.
Our system precisely reproduces the relevant dynamics for para-bosons and para-fermions of even order. 
These results represent the first experimental realization of para-particle dynamics in any physical system.
They also demonstrate full control of para-particle oscillators using a trapped-ion experiment and open the door for the verification of past proposals and future applications for para-particle models.

%%%--------------------------------
\section{Para-particle oscillators}
%%%--------------------------------
A para-particle oscillator may be understood as a parity--deformed oscillator that generalizes the standard Fermi--Dirac and Bose--Einstein statistics associated with fermions and bosons \cite{Macfarlane1994p1054, Dunne1995p3889, Plyushchay1997p619}. 
In the interaction picture, the Hamiltonian for a driven para-particle oscillator,
\begin{eqnarray}\label{eq:pP_Interaction}
	\hat{H}_{\vartheta} = \frac{\hbar g}{2} \left(\hat{A}_{\vartheta} + \hat{A}_{\vartheta}^{\dagger}\right),
\end{eqnarray}
is given in terms of the raising (lowering) operators, $\hat{A}^{\dagger}_{\vartheta}~(\hat{A}_{\vartheta})$, of para-fermions, $\vartheta=pF$, and para-bosons, $\vartheta=pB$, that define a para-particle algebra, see Appendix \ref{app:representation} for more details.
This can be related to an ion system coupled to two motional modes through the effective operators,
\begin{eqnarray}\label{eq:pF_operators}
    \hat{A}_{pF} &=& \sqrt{2} \left( \hat{a}_{x} \hat{\sigma}_{+} + \hat{a}_{y}^{\dagger} \hat{\sigma}_{-} \right),\\ \label{eq:pB_operators}
    \hat{A}_{pB} &=& \sqrt{2} \left( \hat{a}_{x} \hat{\sigma}_{-} - \hat{a}_{y} \hat{\sigma}_{+} \right),
\end{eqnarray}
where $\hat{\sigma}_{+}~(\hat{\sigma}_{-})$ is the spin-raising (lowering) operator and $\hat{a}^{\dagger}_{j}~\left(\hat{a}_{j}\right)$, with $j=x,y$, is the phonon creation (annihilation) operator.
In order to simulate this oscillator, it is necessary to engineer an effective Hamiltonian where both field modes couple simultaneously to the spin under the Jaynes--Cummings (JC) dynamics, Eq. (\ref{eq:pF_operators}), equivalent to a para-Fermi oscillator of even order.
When the $y~(x)$ motional mode is coupled to the spin system under (anti-) JC dynamics, Eq. (\ref{eq:pB_operators}), we create a system that represents a driven para-Bose oscillator, also of even order.
See Appendix \ref{app:mapping} or Refs.  \cite{HuertaAlderete2018p11572,HuertaAlderete2017p013820} for a complete derivation.
%%%-------- Figure ------------------------
\begin{figure}[t]
	\includegraphics[scale=1]{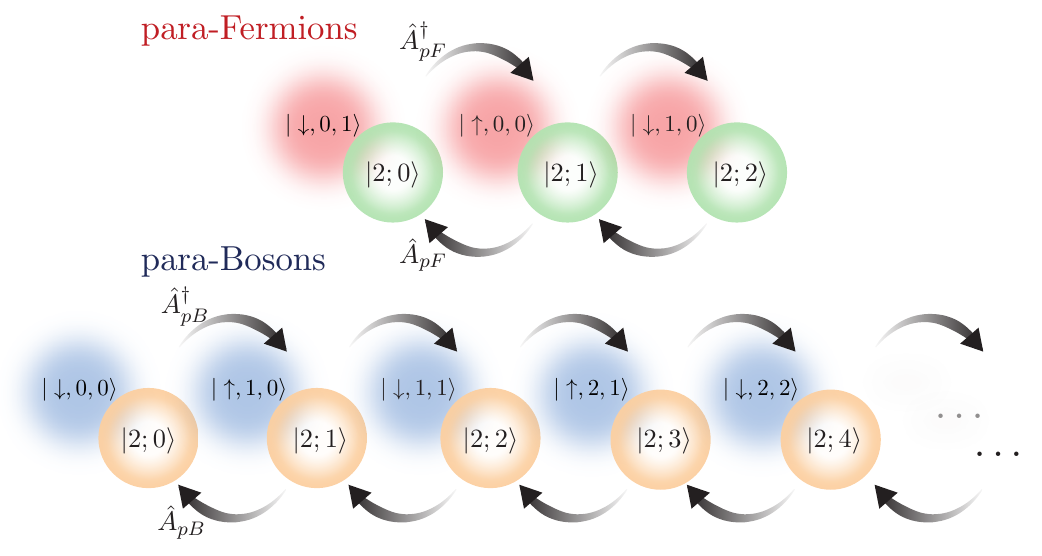}
	\caption{{\bf Ladder of para-particle states.} The states in the ion frame (red and blue shadows), $\vert \psi, n_{x}, n_{y} \rangle$,  correspond to a ladder of a para-particle states (green and orange spheres), $\vert p; k\rangle$, are shown here for order $p=2$. The action of the para-particle raising and lowering operators, $\hat{A}^{\dagger}_{\vartheta}$ and $\hat{A}_{\vartheta}$, are also indicated (arrows) with $\vartheta=pF$ for para-fermions and $\vartheta=pB$ for para-bosons.} \label{fig:Ladder_States}
\end{figure}
%%%-------- Figure ------------------------

The representation implemented in our experiment allows for vacuum states such that $\hat{A}_{\vartheta} \vert p;~\text{vac} \rangle = 0 $. The para-Fermi vacuum state of order $2n$ takes the form $\vert 2n;~ \text{vac}\rangle \equiv \vert \downarrow, 0, n \rangle$, and it is part of a $(2n+1)$-dimensional Hilbert space.
In the para-Bose case of order $2(n+1)$, there can be two vacuum states, $\vert 2(n+1);~\text{vac} \rangle \equiv \vert \downarrow, n, 0 \rangle $ and $\vert 2(n+1);~\text{vac} \rangle \equiv \vert \uparrow, 0, n \rangle $, each part of an infinite-dimensional Hilbert space.
Figure \ref{fig:Ladder_States} illustrates the ladder of para-particle states of order $2$; the states $\left\{\vert \downarrow, 1, 0 \rangle, \vert \uparrow, 0, 0 \rangle, \vert \downarrow, 0, 1 \rangle\right\}$ and $\left\{\vert \downarrow, 0, 0 \rangle, \vert \uparrow, 1, 0 \rangle, \vert \downarrow, 1, 1 \rangle, \ldots\right\}$ form an orthonormal basis for para-fermions and para-bosons, respectively.
By choosing an adequate initial state, we can simulate any para-particle oscillator of even order on a trapped ion system. Measurements of accessible observables in the ion frame, such as the spin $\langle \hat{\sigma}_{z}\rangle$ and the average phonon number $\langle \hat{n}_{j}\rangle$, can be related to physical quantities in the para-particle frame, such as the expectation value of the para-particle number $\langle \hat{\mathcal{N}}_{\vartheta} \rangle$,
\begin{eqnarray} \label{eq:N_pF}
    \langle \hat{\mathcal{N}}_{pF} \rangle &=& \langle \hat{n}_{x} \rangle - \langle\hat{n}_{y} \rangle + \frac{p}{2} , \\ \label{eq:N_pB}
    \langle \hat{\mathcal{N}}_{pB} \rangle &=& \langle \hat{n}_{x} \rangle + \langle\hat{n}_{y} \rangle + 1 - \frac{p}{2},
\end{eqnarray}
where $p$ is the para-particle order.

%%%--------------------------------
\section{Experimental setup}
\begin{figure}[b]
	\includegraphics[scale=1]{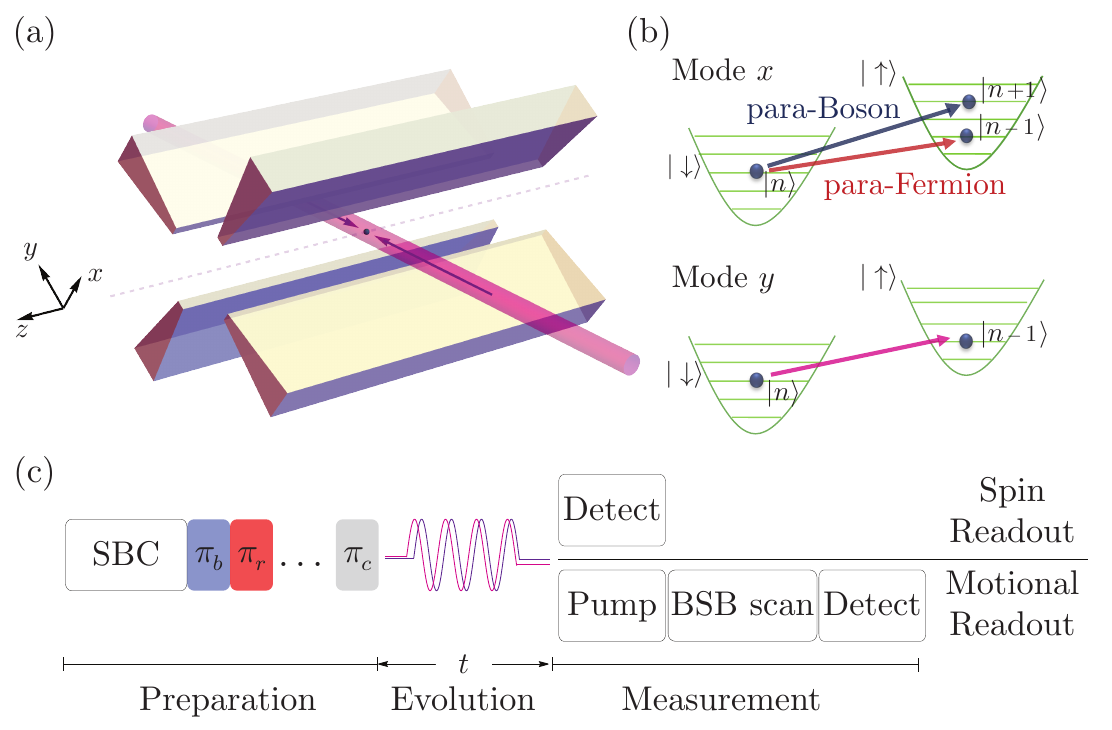}
	\caption{{\bf Experimental scheme.} (a) The system consist of a single $^{171}$Yb$^{+}$ ion confined in a linear Paul trap, addressed by a pair of counter-propagating Raman beams. (b) Applying a drive near the red (blue) sideband of the $x$ mode produces a para-Fermi (para-Bose) interaction in combination with a simultaneous red sideband on the $y$ mode. (c) Experimental sequence consisting of state preparation by sideband cooling and Fock state initialization pulses, para-particle evolution and spin or motional read out.
	}
	\label{fig:pP_Schema}
\end{figure}

%%%--------------------------------
We demonstrate the experimental realization of para-particles using a trapped $^{171}$Yb$^+$ ion confined in a linear Paul trap.
The ion is more tightly confined along the two transverse directions, $x$ and $y$, with mode frequencies  $\omega_{x} = 2\pi \times 3.05$ MHz and $\omega_{y} = 2\pi \times 2.88$ MHz, respectively, which are used as the bosonic modes for the simulation, see Fig. \ref{fig:pP_Schema}(a).
%%%-------- Figure ------------------------
The ion is laser-cooled close to the motional ground state of both of these modes by Doppler cooling and subsequent Raman sideband cooling.
The spin-$1/2$ system is encoded in the hyperfine-split $^{2}S_{1/2}$ manifold, where we choose $\vert \downarrow \rangle \equiv \vert F=0; m_{F}=0 \rangle$ and $\vert \uparrow \rangle \equiv \vert F=1; m_{F}=0 \rangle$ with a frequency difference $\omega_{HF}=2\pi \times 12.642821$ GHz, which is insensitive to magnetic field fluctuations to first order.
The spin is initialized by optical pumping and read out using state-dependent fluorescence detection \cite{Olmschenk2007p052314}.
A pair of counter-propagating Raman beams from a single $355$-nm mode-locked laser \cite{Islam2014p3238} are used to manipulate the spin of the ion.
These beams can also impart momentum on the ion to change the motional state in the trap by addressing the motional sidebands $\omega_{HF} \pm \omega_{x(y)}$. In the interaction picture, those Raman laser operations can be described by the following Hamiltonians,
\begin{eqnarray}\label{eq:rsb}
	\hat{H}_{JC} &=& \frac{\hbar \eta \Omega_r}{2} \left( \hat{a} \hat{\sigma}_{+} e^{i \phi } + \hat{a}^{\dagger} \hat{\sigma}_{-} e^{-i \phi}\right),\\ \label{eq:bsb}
	\hat{H}_{aJC} &=& \frac{\hbar \eta \Omega_b}{2} \left( \hat{a} \hat{\sigma}_{-} e^{i \phi } + \hat{a}^{\dagger} \hat{\sigma}_{+} e^{-i \phi}\right).
\end{eqnarray}
Here again, $\hat{\sigma}_{+}~(\hat{\sigma}_{-})$ is the spin-raising (lowering) operator, $\hat{a}^\dagger~(\hat{a})$ is the phonon creation (annihilation) operator, $\Omega_{r(b)}$ is the sideband Rabi frequency of the (anti-) JC model or red (blue) sideband, $\eta= \Delta k \sqrt{\hbar/M \omega_{j}}$ is the Lamb-Dicke parameter where $\Delta k$ is the net wave vector of the Raman laser beams, $M$ is the mass of the $^{171}$Yb$^+$ ion, and $\phi$ is the phase difference between the Raman laser beams.
%Each sideband is generated by driving resonant Rabi rotations of defined phase, amplitude, and duration \cite{Islam2014p3238}.
During the experiment, we drive a combination of those operations, Eqs. (\ref{eq:rsb})-(\ref{eq:bsb}), on both modes, $x$ and $y$, simultaneously, see Fig. \ref{fig:pP_Schema}(b), to realize the dynamics of a para-particle oscillator \cite{HuertaAlderete2017p013820,HuertaAlderete2018p11572}.
A typical experimental sequence is shown in Fig. \ref{fig:pP_Schema}(c). 
First, we perform cooling of the motion and optical pumping to prepare the system in the state $\vert \downarrow, 0, 0 \rangle$ \cite{Monroe1995p4011}, and then transfer the system to a specific initial state $\vert \psi, n_{x}, n_{y} \rangle$ based on the choice of para-particle order.
To initialize the motional modes in a particular Fock state, phonon excitations are introduced by resonantly driving a combination of blue-sideband, red-sideband and carrier $\pi$-pulse to prepare the desired state, see Appendix \ref{App:FSPrep} for more details. 
Then, we off-resonantly drive the two sideband transitions responsible for the para-particle dynamics with precisely defined intensities and detunings. 
After the para-particle evolution, we read out either the spin or the motional state.
While the spin measurement is made directly with the expectation value calculated by averaging over many repetitions, the motional readout requires the spin as an intermediary.
First, we reset the spin to $\vert \downarrow \rangle$ by applying an optical pumping pulse with negligible effect on the motional state population since our system is in the Lamb-Dicke regime.
Subsequently, a blue-sideband pulse with frequency $\omega_{x(y)}$ is applied to probe the Fock state distribution.
This is repeated with increasing pulse duration resulting in Rabi oscillations with frequency components whose amplitudes depend on the Fock state populations \cite{Meekhof1996p1796}.
The resulting signal is fitted to a sum of sinusoids to obtain their relative amplitudes, with the frequencies and decay rates determined by an independent measurement of the Rabi frequency, more details are given in Appendix \ref{App:MotionalAnalysis}.

%%%--------------------------------
\section{Results}
%%%--------------------------------
\subsection{Para-fermions}
\begin{figure}[h]
	\includegraphics[scale=1]{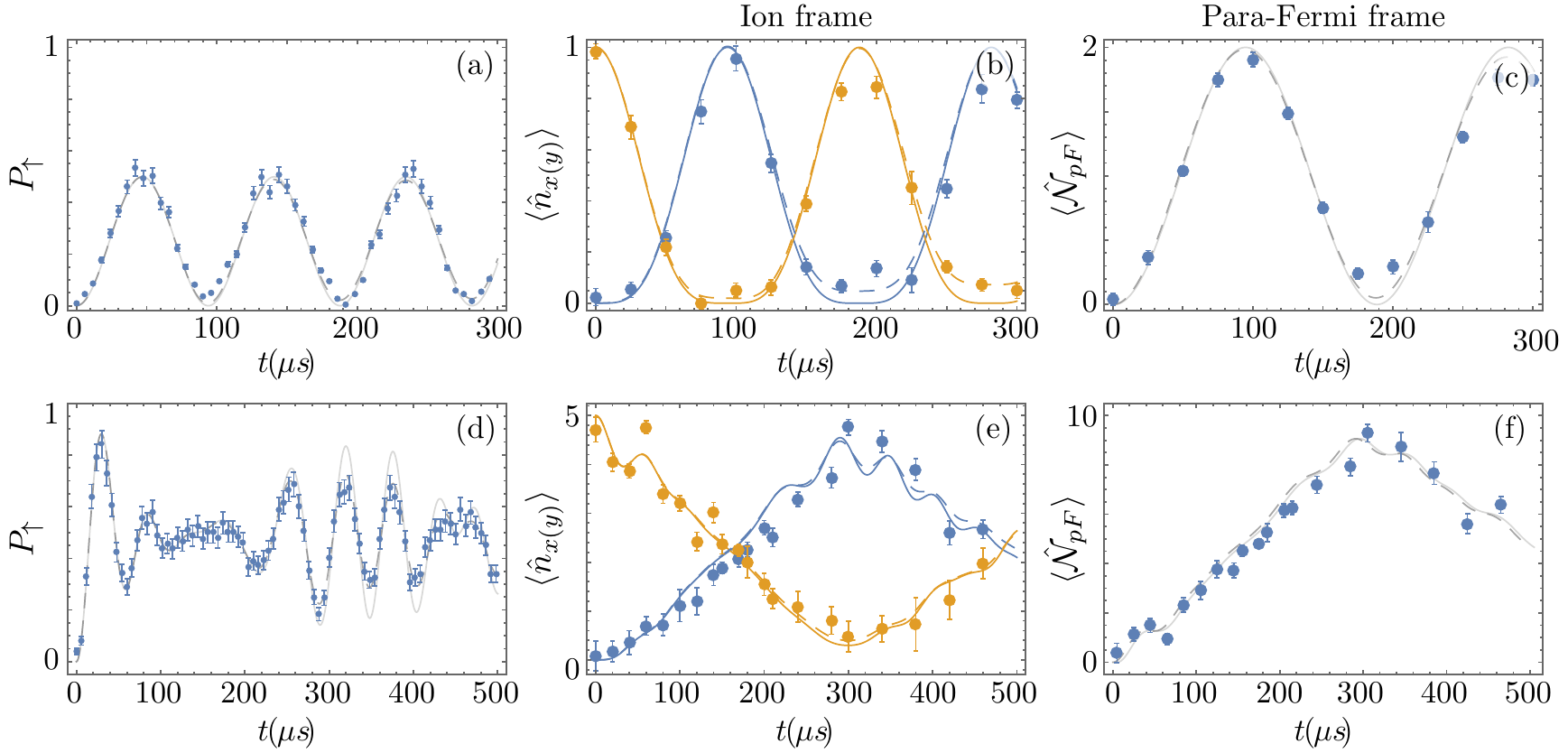}
	\caption{{\bf Para-Fermi dynamics.} Experimental realization of a para-Fermi oscillator of order 2 (upper row) and 10 (lower row) starting from the vacuum state. (a) and (d) show the spin evolution, $P_{\uparrow}$, (b) and (e) the evolution of the average phonon number, $\langle \hat{n}_{x(y)} \rangle$, in the $x~(y)$ mode, blue (yellow) line, (c) and (f) the corresponding evolution of the para-Fermi number operator, $\langle \hat{\mathcal{N}}_{pF} \rangle$. Continuous lines are simulations of the dynamics based on our experimental settings with no free parameters, dashed lines include effects of motional heating.}
	\label{fig:parafermions}
\end{figure}

%%%--------------------------------
We first simulate a para-fermion of order $2$, hence we prepare the vacuum state $\vert 2; 0 \rangle \equiv \vert \downarrow, 0,1 \rangle$.
The simulation consists of driving the red sidebands of both motional modes with equal Rabi frequency, Eq. (\ref{eq:pF_operators}).
The spin oscillates with amplitude $0.5$, Fig. \ref{fig:parafermions}(a), and facilitates excitation exchange between $x$ (blue line) and $y$ (yellow line) motional modes, Fig. \ref{fig:parafermions}(b). After a full oscillation of the spin, the population of the $y$ mode switches coherently to the $x$ mode. This represents the evolution of the para-Fermi number operator, $\langle \hat{\mathcal{N}}_{pF} \rangle$ in Eq. (\ref{eq:N_pF}), in the para-Fermi frame, shown in Fig. \ref{fig:parafermions}(c).
The dynamics become more complex as we increase the order, and hence scale up the number of occupied Fock states in the trapped-ion system.
We also realize a system of order $10$, starting from the vacuum state $\vert 10; 0\rangle \equiv\vert \downarrow, 0,5 \rangle$.
The spin evolution exhibits oscillations that collapse and then revive partially, Fig. \ref{fig:parafermions}(d), a feature also seen in the excitation exchange between motional modes, Fig. \ref{fig:parafermions}(e).
This translates into the expected para-Fermi number excitation, Fig. \ref{fig:parafermions}(f).
%%%-------- Figure ------------------------

The data points in Figure \ref{fig:parafermions} correspond to experimental results.
Every data point is acquired by averaging over 300 measurements of the experimental sequence, and the error bars are statistical for the spin population and a result of the fitting procedure for the motional modes.
Continuous lines correspond to a numerical simulation with up to eight Fock states per motional mode.
We investigate effects of motional heating in the trap as a potential source of error by including a Lindblad term in our numerical simulation, see Appendix \ref{app:Lindblad}. We find them to be small, especially for shorter times of the evolution, see Fig. \ref{fig:parafermions}.
We attribute the additional deviations visible in some places to fluctuations in the experimental parameters, especially the intensity, which is subject to beam pointing drift.
We note that the evolution of the lowest energy state under driven para-Fermi oscillator dynamics resembles a binomial state in the ion frame \cite{HuertaAlderete2018p11572}.
This produces single frequency oscillations of the spin for small para-particle order, Fig. \ref{fig:parafermions}(a), and a multiple-frequency beat as the order increases, Fig. \ref{fig:parafermions}(d).
To the best of our knowledge, this is the first time that binomial states have been created in a physical system since their theoretical introduction \cite{Stoler1985p345,VidellaBarranco1994p5233}.

%%%--------------------------------
\subsection{Para-bosons}
\begin{figure}[h!]
	\centering
	\includegraphics[scale=1]{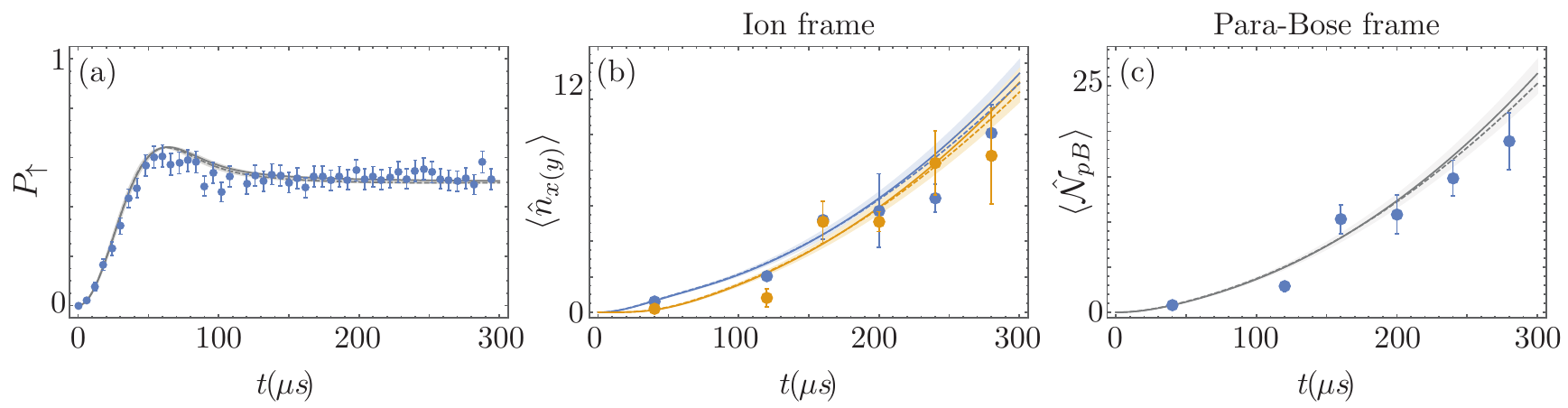}
	\caption{{\bf Para-Bose dynamics.} Experimental realization of a para-Bose oscillator of order 2 starting from the vacuum state. (a) shows the spin evolution, $P_{\uparrow}$, (b) the evolution of the average phonon number, $\langle \hat{n}_{x(y)} \rangle$, in the $x~(y)$ mode, blue (yellow) line, (c) the corresponding evolution of the para-Bose number operator, $\langle \hat{\mathcal{N}}_{pB} \rangle$. Shaded lines are simulations of the dynamics based on our experimental settings with no free parameters, dashed lines include effects of motional heating.}
	\label{fig:pb_p2}
\end{figure}

%%%--------------------------------
We simulate the para-Bose oscillator of order $2$ by first preparing the vacuum state, $\vert 2;0\rangle \equiv \vert \downarrow, 0, 0 \rangle$. We choose a low order due to the fast increase of the Fock state occupation from the simulation dynamics.
The simulation consists of driving near the blue sideband on the $x$ motional mode and the red sideband on the $y$ motional mode with equal Rabi frequencies. The spin evolution corresponds to a damped oscillation, Fig. \ref{fig:pb_p2}(a), while the motional excitations increase, Fig. \ref{fig:pb_p2}(b), reproducing the expected dynamics of the para-Bose number operator,  Fig. \ref{fig:pb_p2}(c).

The experimental outcome is directly compared to a numerical simulation, continuous lines, with a Hilbert space of up to 35 Fock states per motional mode; motional heating is investigated by including a Lindblad term on the simulations, see Appendix \ref{app:Lindblad}.
The para-Bose simulation is more sensitive to anisotropy in the red and blue sideband driving \cite{HuertaAlderete2017p013820}, so we take as an upper(lower) bound the frequency $g_{\pm} = \Omega_{pB} \pm \delta$ with $\Omega_{pB}= (\Omega_{r}+\Omega_{b})/2$ and $\delta= \vert \Omega_{r} - \Omega_{b}\vert/2$, where $\Omega_{r(b)}$ comes from calibration measurements on the experiment. The shadings in Fig. \ref{fig:pb_p2} correspond to these bounds.
As in the para-Fermi case, every data point is acquired by averaging over 300 measurements of the experimental sequence and error bars are statistical for spin population and a result of the fitting procedure for the motional modes.
We note that this time evolution is related to a coherent state \cite{HuertaAlderete2017p013820}, with interesting non-classical properties; the nature of its statistics (sub/super- Poissonian) is controlled by the Rabi frequency of the drive and the para-particle order \cite{HuertaAlderete2017p043835}.
%%%-------- Figure ------------------------
%%%--------------------------------
\section{Conclusion}
%%%--------------------------------
Driving dynamics with trapped ions that involve the Hilbert spaces of multiple orthogonal vibrational modes simultaneously is a promising experimental capability for the simulation of physical systems \cite{Davoudi2021p, Chen2021p060311}.
We demonstrate highly controllable experimental realizations of both kinds of para-particle oscillators for the first time.
Our work is a first step towards the more detailed studies of para-particles that involve single and many body interactions, as well as the search for new properties such as topological phases \cite{Cai2020pnwaa196} or multi-frequency conversion \cite{Kockum2017p2045}.

\begin{acknowledgments}
We thank Elijah Keene Kaimiola Mossman for his help with the experiment.
This work is supported by the National Science Foundation via the Physics Frontier Center at the Joint Quantum Institute (PHY-1430094), the Maryland-ARL Quantum Partnership (W911NF1920181), and the DOE Office of Science, Office of Nuclear Physics (DE-SC0021143). A. M. G. is supported by a Joint Quantum Institute Postdoctoral Fellowship.
\end{acknowledgments}
%%%%%%%%%% Supplemental materials %%%%%%%%%%
\nocite{*}

%\newpage
% \onecolumngrid
\appendix

%%%%%%%%% Merge with supplemental materials %%%%%%%%%%
% \begin{center}
% \textbf{\large Supplemental Materials: First experimental realization of para-particle oscillators}
% \end{center}
%%%%%%%%% Merge with supplemental materials %%%%%%%%%%
%%%%%%%%% Prefix a "S" to all equations, figures, tables and reset the counter %%%%%%%%%%
% \setcounter{equation}{0}
% \setcounter{section}{0}
% \setcounter{figure}{0}
% \setcounter{table}{0}
% \setcounter{page}{1}
% \makeatletter
% \renewcommand{\theequation}{S\arabic{equation}}
% \renewcommand{\thefigure}{S\arabic{figure}}
% \renewcommand{\bibnumfmt}[1]{[S#1]}
% \renewcommand{\citenumfont}[1]{S#1}
%%%%%%%%%% Prefix a "S" to all equations, figures, tables and reset the counter %%%%%%%%%%

%------------------------------------------
\section{Para-particle representation}\label{app:representation}
%------------------------------------------
Para-particle creation (annihilation) operators $\hat{A}_{\vartheta}~(\hat{A}^{\dagger}_{\vartheta})$ follow the so-called tri-linear commutation relations,
\begin{eqnarray}
    \left[ \left[ \hat{A}_{pF}, \hat{A}^{\dagger}_{pF} \right] , \hat{A}_{pF} \right] = -2\hat{A}_{pF}, &\qquad& \left[ \left[ \hat{A}_{pF}, \hat{A}^{\dagger}_{pF} \right] , \hat{A}_{pF}^{\dagger} \right] = 2\hat{A}_{pF}^{\dagger},\\
    \left[ \left\{ \hat{A}_{pB}, \hat{A}^{\dagger}_{pB} \right\} , \hat{A}_{pB} \right] = -2\hat{A}_{pB}, &\qquad& \left[ \left\{ \hat{A}_{pB}, \hat{A}^{\dagger}_{pB} \right\} , \hat{A}_{pB}^{\dagger} \right] = 2\hat{A}_{pB}^{\dagger}.
\end{eqnarray}
These relations include the usual bosons and fermions. We consider a driven oscillator Hamiltonian,
\begin{eqnarray}
    \hat{H}_{d} &=& \hat{H}_{0} + \hbar g \left( \hat{A}_{\vartheta} + \hat{A}_{\vartheta}^{\dagger}\right) \cos \omega_{d}t,
\end{eqnarray}
where $\hat{H}_{0}$ is the free para-particle Hamiltonian defined as $\hat{H}_{0} =\frac{\hbar \omega}{2}\left\{ \hat{A}_{pB}, \hat{A}_{pB}^{\dagger}\right\}$ for para-bosons and  $\hat{H}_{0} =\frac{\hbar\omega}{2}\left[ \hat{A}_{pF}, \hat{A}_{pF}^{\dagger}\right]$ for para-fermions. 
We move to the frame given by the free para-particle Hamiltonian to obtain,
\begin{eqnarray}
    \hat{H}_{\vartheta} &=& e^{i \hat{H}_{0}t/\hbar} \hat{H}_{d} e^{-i \hat{H}_{0}t/\hbar} \nonumber \\
    &=& \hbar g \left( \hat{A}_{\vartheta} e^{-i\omega t} + \hat{A}_{\vartheta}^{\dagger}  e^{i\omega t}\right) \cos \omega_{d}t.
\end{eqnarray}
Assuming resonant driving, $\vert \omega - \omega_{d}\vert = 0$, and weak coupling, $\vert \omega + \omega_{d}\vert \gg g$, allows us to apply a rotating wave approximation to simplify this interaction Hamiltonian, 
\begin{eqnarray} \nonumber
    \hat{H}_{\vartheta} &=& \frac{\hbar g}{2} \left( \hat{A}_{\vartheta} + \hat{A}_{\vartheta}^{\dagger} \right).
\end{eqnarray}
We use a parity-deformed representation \cite{Plyushchay1997p619},
\begin{eqnarray} \label{eq:pF_commutations}
    \left\{ \hat{A}^{\dagger}_{pF}, \hat{A}_{pF} \right\}  &=&  (p+1) -\hat{\mathcal{R}}_{pF},  \qquad  \left[ \hat{A}^{\dagger}_{pF}, \hat{A}_{pF} \right] = 2\left( \hat{\mathcal{N}}_{pF} -\frac{p}{2} \right)\hat{\mathcal{R}}_{pF}, \\ 
    \left[ \hat{A}_{pB}, \hat{A}^{\dagger}_{pB}\right] &=& 1 + (p-1)\hat{\mathcal{R}}_{pB}, \qquad \frac{1}{2}\left\{ \hat{A}_{pB}, \hat{A}^{\dagger}_{pB}\right\} = \hat{\mathcal{N}}_{pB} + \frac{p}{2},\label{eq:pB_commutations} \\\nonumber
    &&\left\{ \hat{\mathcal{R}}_{\vartheta}, \hat{A}_{\vartheta} \right\} = \left\{ \hat{\mathcal{R}}_{\vartheta}, \hat{A}^{\dagger}_{\vartheta} \right\} = 0.
\end{eqnarray}% 
where $\hat{\mathcal{R}}_{\vartheta} = e^{-i \pi \hat{\mathcal{N}}_{\vartheta}}
$ is the parity operator, such that $\hat{\mathcal{R}}_{\vartheta}^{2} = 1$, and $\hat{\mathcal{N}}_{\vartheta}$ is the para-particle number operator. For a given order we can define an orthonormal Hilbert space where the para-particle order is given by the action,
\begin{eqnarray}
    \hat{A}_{\vartheta} \hat{A}^{\dagger}_{\vartheta} \vert p; \text{vac} \rangle = p \vert p; \text{vac} \rangle,
\end{eqnarray}
 on the vacuum state.
%The representation used for para-Fermions holds a slightly different tri-linear relation, $\left[ \left[ \hat{A}_{pF}, \hat{A}^{\dagger}_{pF} \right] , \hat{A}_{pF} \right] = 2\left(2\hat{A}_{0} +1\right)\hat{\mathcal{R}}\hat{A}_{pF}$, and $\left[ \left[ \hat{A}_{pF}, \hat{A}^{\dagger}_{pF} \right] , \hat{A}_{pF}^{\dagger} \right] = 2\left(2\hat{A}_{0} - 1\right)\hat{\mathcal{R}}\hat{A}_{pF}^{\dagger}$, the shape of the Hamiltonian of interaction, \ref{eq:pP_Interaction}, holds when working under the rotating wave approximation and resonance conditions.

%%%--------------------------------
\section{Mapping}\label{app:mapping}
%----------------------------------
Our simulation starts with the interaction Hamiltonian in the ion frame,
\begin{eqnarray}
    \hat{H}_{ion}^{pF} &=&  \frac{\hbar\Omega}{2} \left[\left( \hat{a}_{x} + \hat{a}_{y} \right) \hat{\sigma}_{+} + \left( \hat{a}_{x}^{\dagger} + \hat{a}_{y}^{\dagger} \right) \hat{\sigma}_{-} \right], \\
    \hat{H}_{ion}^{pB} &=&  \frac{\hbar\Omega}{2}\left[\left( \hat{a}_{x}^{\dagger} - \hat{a}_{y} \right) \hat{\sigma}_{+} + \left( \hat{a}_{x} - \hat{a}_{y}^{\dagger} \right) \hat{\sigma}_{-} \right],
\end{eqnarray}
where $\Omega=\Omega_{r}=\Omega_{b}$ is determined by experimental parameters. Then, we identify the para-particle operators, Eqs. (\ref{eq:pF_operators})-(\ref{eq:pB_operators}), such that we recover the interaction Hamiltonian for a driven para-particle oscillator, Eq. (\ref{eq:pP_Interaction}). Table \ref{tab:pP} summarizes the relations between the two frames.  
\begin{table}[h!]
\centering
    \begin{tabular}{c c cc cc c}
    & &&  && & \\ \hline\hline
    & &&  && & \\
    & Para-particle frame & \multicolumn{4}{c}{Ion-frame} &\\
    & &&  && &  \\ \hline\hline
    & &&  && &  \\ 
    &&& para-fermions && para-bosons &\\  
    & &&  && & \\    \hline   \hline  
    %  &&  &&  \\                  
    % $\hat{\mathcal{O}}_{\vartheta}$ &\qquad \qquad& $\hat{n}_{x} + \hat{n}_{y} + \frac{1}{2}\left(1+\hat{\sigma}_{z} \right)$ &\qquad \qquad& $\hat{n}_{y} - \hat{n}_{x} + \frac{1}{2}\left(1+\hat{\sigma}_{z} \right)$ \\
    & &&  && &  \\        
    &$\hat{A}_{\vartheta}$ && $\sqrt{2} \left( \hat{a}_{x} \hat{\sigma}_{+} + \hat{a}_{y}^{\dagger} \hat{\sigma}_{-} \right)$ && $\sqrt{2} \left( \hat{a}_{x} \hat{\sigma}_{-} - \hat{a}_{y} \hat{\sigma}_{+} \right)$ &\\
    & &&  && & \\        
   & $\left[ \hat{A}_{\vartheta}, \hat{A}^{\dagger}_{\vartheta}\right] $ && $2 \left(\hat{n}_{x} - \hat{n}_{y} \right)\hat{\sigma}_{z}$ && $2 \left(\hat{n}_{y} - \hat{n}_{x}\right) \hat{\sigma}_{z} + 2 $ &\\
    & &&  && & \\           
    &$\left\{ \hat{A}^{\dagger}_{\vartheta}, \hat{A}_{\vartheta}\right\} $ && $ 2 \left(\hat{n}_{x} + \hat{n}_{y}+ 1 + \hat{\sigma}_{z}\right)$ && $2 \left(\hat{n}_{x} + \hat{n}_{y} + 1 \right) $& \\
    & &&  && & \\            
    &Parity && $-\hat{\sigma}_{z}$ && $\hat{\sigma}_{z}$&\\
    & &&  &&  &\\       
    &$\hat{\mathcal{N}}_{\vartheta}$ && $\hat{n}_{x} - \hat{n}_{y} + \frac{p}{2}$ &&  $\hat{n}_{x} + \hat{n}_{y} + 1 - \frac{p}{2}$ &\\
    & &&  &&  &\\      
    % $p$ && $ \langle\hat{\mathcal{O}}_{pF}\rangle+1$ &&  $ 2\left(\langle\hat{\mathcal{O}}_{pB}\rangle+1\right)$ \\
    %  &&  &&  \\      
    \hline
    \end{tabular}
\caption{Relations between operators para-particle frame and the ion.}
\label{tab:pP}
\end{table}

%%%--------------------------------
\section{Fock state preparation}\label{App:FSPrep}
\begin{figure}[h]
	\includegraphics[scale=1]{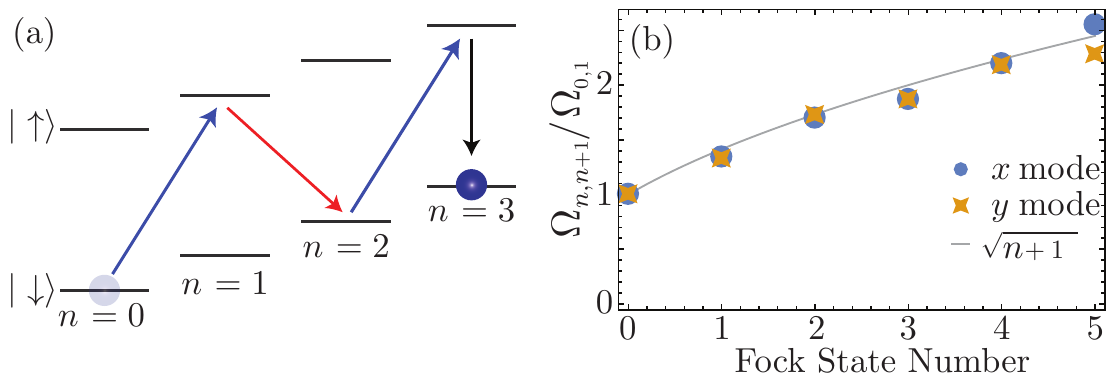}
	\caption{{\bf Fock State Preparation} (a) Sequence of red sideband, blue sideband and carrier pulses to prepare $\vert \downarrow, n=3 \rangle$. (b) the measured relative Rabi frequencies $\Omega_{n,n+1} /\Omega_{0,1}$. The solid line corresponds $\Omega_{n,n+1} /\Omega_{0,1}=\sqrt{n+1}$, which holds in the Lamb-Dicke regime.}
	\label{fig:App_FSprep}
\end{figure}

%----------------------------------
The blue sideband, or anti-Jaynes--Cummings, interaction induces transitions between the states $\vert \downarrow , n \rangle$ and $\vert \uparrow , n + 1 \rangle$ with Rabi frequencies proportional to $\sqrt{n+1}$.
The red sideband, or Jaynes--Cummings, interaction drives transitions between the states $\vert \downarrow, n \rangle$ and $\vert \uparrow, n - 1 \rangle$ with Rabi frequencies proportional to $\sqrt{n}$.
The carrier transition changes the state of the spin without changing the Fock state. 
Our system is initially cooled to the ground state $\vert \downarrow, 0\rangle$.
We use a series of sideband and carrier $\pi$-pulses to initialize different Fock states, Fig. \ref{fig:App_FSprep}(a).
For example, to prepare the Fock state $n=3$, the pulse sequence we need is blue sideband--red sideband--blue sideband--carrier, which will walk the excitation through the states $\vert \downarrow, 0 \rangle \rightarrow \vert \uparrow, 1\rangle \rightarrow \vert \downarrow, 2 \rangle \rightarrow \vert \uparrow, 3\rangle  \rightarrow \vert \downarrow, 3\rangle$. 

Each measured probability for $\vert \downarrow, n\rangle$ with $n=0,1,\ldots,5$ is fitted to $P_{\uparrow}(t) = e^{-\gamma_{n} t}\cos^{2} \Omega_{n,n+1} t$, where $\Omega_{n,n+1}$ is the Rabi frequency and $\gamma_{n}$ is the decoherence rate between levels $\vert n\rangle$ and $\vert n+1 \rangle$.
The measured Rabi frequency ratios $\Omega_{n,n+1} /\Omega_{0,1}$ are plotted in Figure \ref{fig:App_FSprep}(b), and compared to the expected $\sqrt{n+1}$.
We observe no inherent cross-coupling between excitations on $x$ and $y$ modes. 

%%%--------------------------------
\section{Fock state population measurement}\label{App:MotionalAnalysis}
%%%-------- Figure ------------------------
\begin{figure}[t]
	\includegraphics[scale=1]{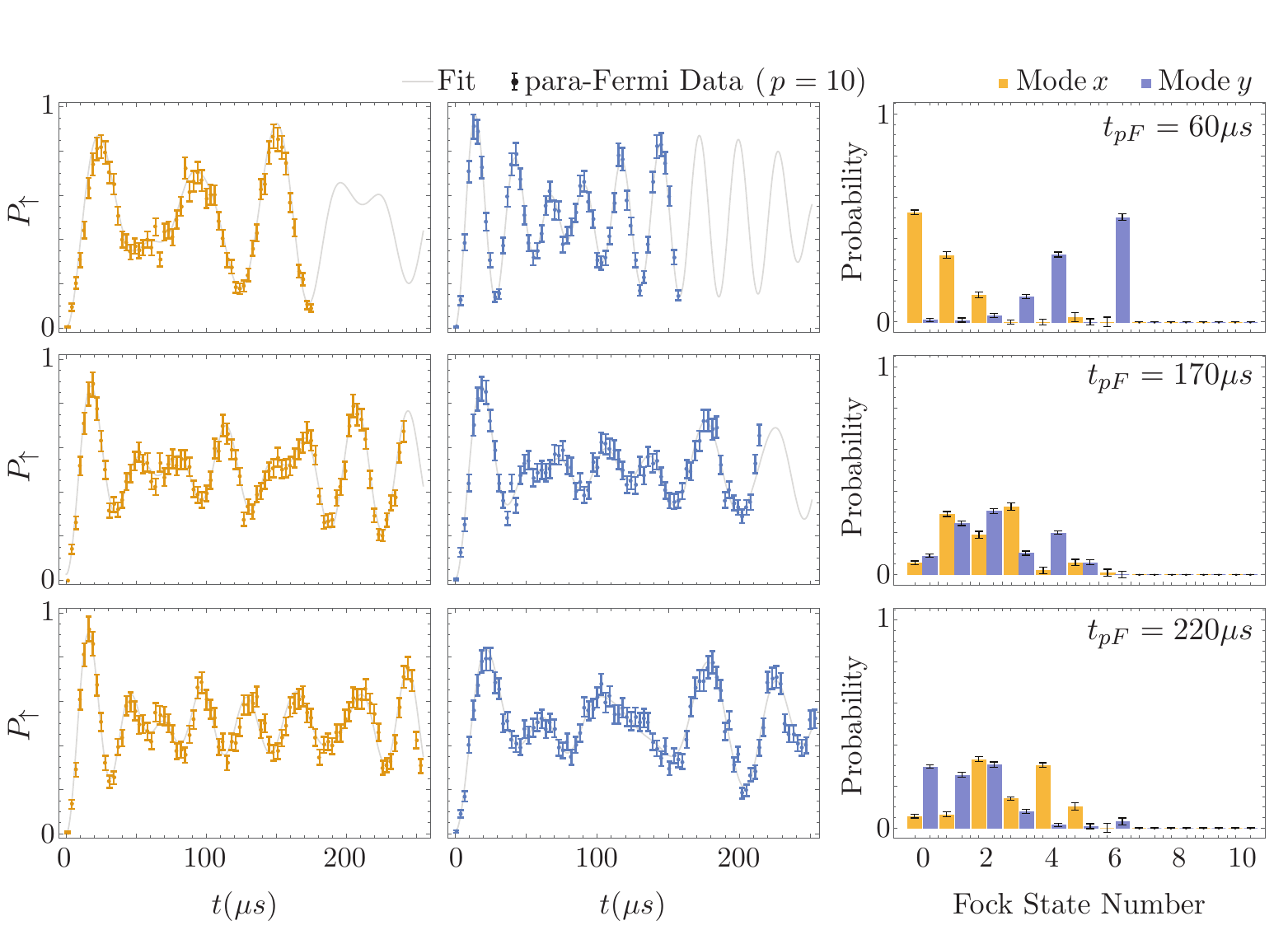}
	\caption{{\bf Fock state population analysis.} Each row corresponds to a particular point in the time evolution of the para-Fermi dynamics of order 10, $t=60\mu s$, $t=170\mu s$, and $t=220 \mu s$, see Fig. \ref{fig:parafermions} in the main text. The first two columns show the blue sideband evolution in the $x$ and $y$ modes, respectively. Continuous lines correspond to fits of the experimental data. The right-most column shows the resulting Fock state distributions for two modes.
	}
	\label{fig:pF_detection}
\end{figure}
\begin{figure}[h]
	\includegraphics[scale=1]{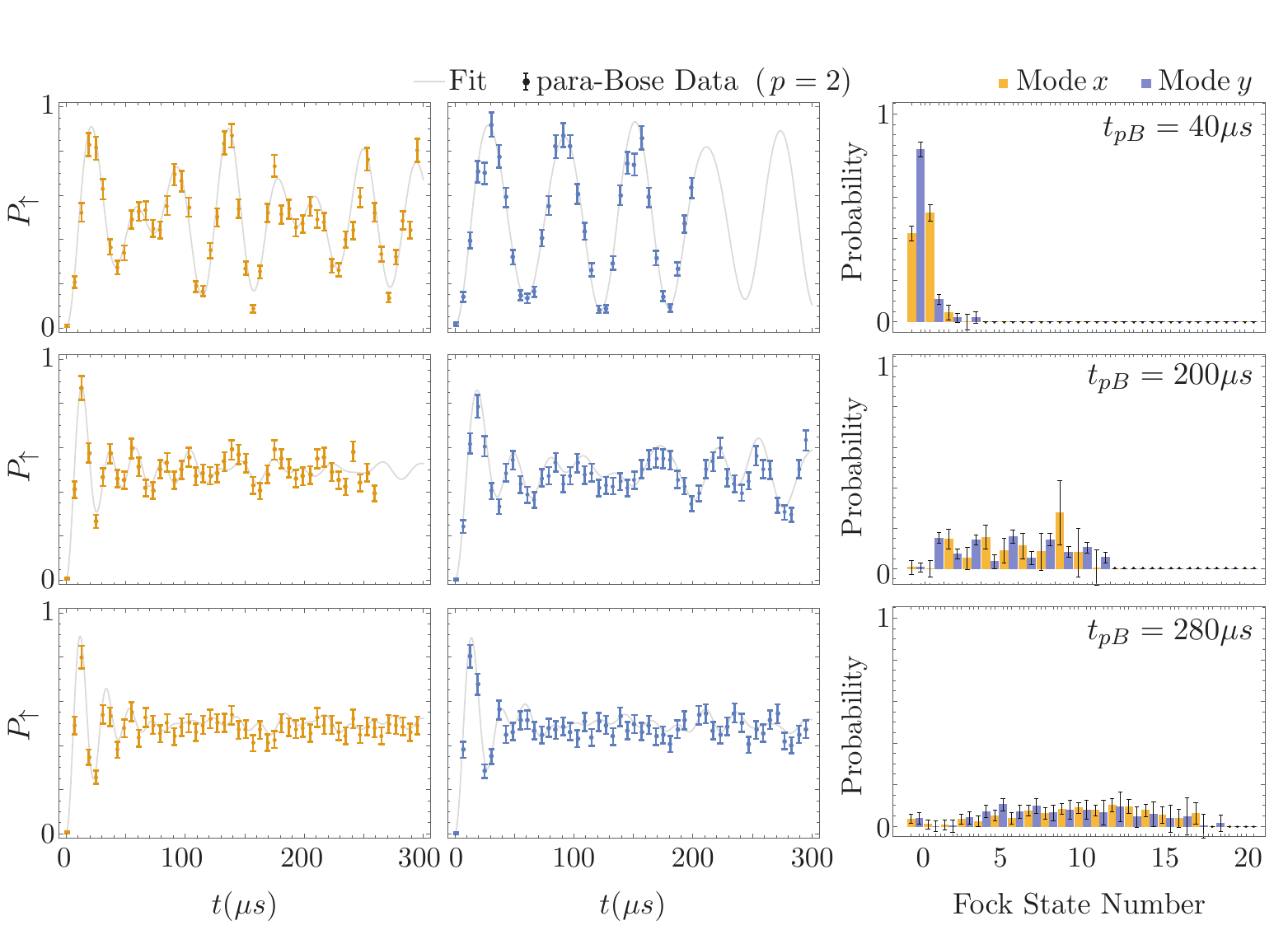}
	\caption{{\bf Fock state population analysis.} Each row shows a particular point in the time evolution of the para-Bose dynamics of order $2$, $t=40\mu s$, $t=200\mu s$, and $t=280 \mu s$, see Fig. \ref{fig:pb_p2} in the main text. The first two columns show the blue sideband evolution in the $x$ and $y$ modes, respectively. Continuous lines correspond to fits of the experimental data. The right-most column shows the resulting Fock state distribution for the two modes.}
	\label{fig:pB_detection}
\end{figure}

Fock state populations are analyzed by first setting the spin to $\vert \downarrow \rangle$ by optical pumping and then applying a blue sideband pulse for various interaction times before reading out the spin state \cite{Meekhof1996p1796}.  The resulting curve is described by the function,
\begin{eqnarray}\label{eq:Prob_down}
	P_{\uparrow}(t) = \frac{1}{2} \left(1 + \sum_{n=0}^{\infty} P_{n} e^{-\gamma_{n}t} \cos \Omega_{n,n+1}t\right),
\end{eqnarray}
from which the desired Fock state populations $P_{n}$ can be obtained by a fit. The $\gamma_{n}$ are phenomenological decay constants assumed to obey $\gamma_{n}= \gamma \sqrt{n+1}$. Notice that the more populated the motional Hilbert space, the greater the number of frequency components required for the fitting procedure. Typical examples of this measurement and fitting procedure are shown in Fig. \ref{fig:pF_detection} for para-Fermi and Fig. \ref{fig:pB_detection} for para-Bose at various times.% From the measured phonon distribution, $P_{\uparrow}(t)$, we can then calculate the average phonon number for each mode and the para-particle free energy for each para-oscillator.

%-----------------------------------
\section{Numerical Simulations}\label{app:Lindblad}
%-----------------------------------
We investigate effects of motional heating by computing the time evolution of the system according to a Lindblad master equation,
\begin{eqnarray}
    \partial_{t} \hat{\rho} = -i \left[ \hat{H}^{\vartheta}_{ion}, \hat{\rho}\right] + \sum_{j=x,y}\left( \gamma_{j} n_{th} \hat{\mathcal{L}}[\hat{a}^{\dagger}_{j}]\hat{\rho} +  \gamma_{j}(n_{th}
    +1) \hat{\mathcal{L}}[\hat{a}_{j}]\hat{\rho}\right),
\end{eqnarray}
where,
\begin{eqnarray}
     \hat{\mathcal{L}}[\hat{O}]\hat{\rho} = 2 \hat{O} \hat{\rho} \hat{O}^{\dagger} - \hat{O}^{\dagger}\hat{O}\hat{\rho} - \hat{\rho} \hat{O}^{\dagger} \hat{O}
\end{eqnarray}
is the Lindblad operator and $\gamma n_{th} \approx \gamma(n_{th}+1)$ is the motional heating rate which is measured as $70$ phonons$/s$ in our system. 

%%%-------- Figure ------------------------
%\twocolumngrid
\bibliography{parafermions}
\end{document}